\documentclass[fleqn, usenatbib]{mnras}
\usepackage{newtxtext,newtxmath}
\usepackage[T1]{fontenc}
\usepackage{ae,aecompl}
\usepackage[]{units}
\usepackage{graphicx}   
\usepackage{amsmath}    
\usepackage{amssymb}    
\usepackage{graphicx}   
\usepackage{amsmath}    
\usepackage{amssymb}    
\usepackage{layouts}
\usepackage{multicol}        
\usepackage{bm}         
\usepackage{pdflscape}  
\usepackage{verbatim}

\def\msun{M$_{\sun}$}

\def\aap{A\&A}
\def\apjl{ApJ}
\def\apj{ApJ}
\def\apjs{ApJS}

\def\mnras{MNRAS}

\def\pasp{PASP}

\def\araa{ARAA}

\title[A search for pulsations in MSP companions]
{A Refined Search for Pulsations in White Dwarf Companions to Millisecond Pulsars\thanks{
Based on observations obtained at the Gemini Observatory, which is operated by the 
Association of Universities for Research in Astronomy, Inc., under a cooperative agreement 
with the NSF on behalf of the Gemini partnership: the National Science Foundation 
(United States), the National Research Council (Canada), CONICYT (Chile), the Australian 
Research Council (Australia), Minist\'{e}rio da Ci\^{e}ncia, Tecnologia e Inova\c{c}\~{a}o 
(Brazil) and Ministerio de Ciencia, Tecnolog\'{i}a e Innovaci\'{o}n Productiva (Argentina).}}
\author[M. Kilic et al.]
{Mukremin Kilic$^{1,2}$,
J. J. Hermes$^3\thanks{Hubble Fellow}$,
A. H. C\'orsico$^4$,
Alekzander Kosakowski$^{1}$,
\newauthor 
Warren R. Brown$^5$,
John Antoniadis$^6$,
Leila M. Calcaferro$^4$,
A. Gianninas$^1$,
\newauthor
Leandro G. Althaus$^4$,
M. J. Green$^7$
\\
$^1$Department of Physics and Astronomy, University of Oklahoma, 440 W. Brooks St., Norman, OK, 73019, USA\\
$^2$Institute for Astronomy, University of Edinburgh, Royal Observatory, Blackford Hill, Edinburgh EH9 3HJ, UK\\
$^3$Department of Physics and Astronomy, University of North Carolina, Chapel Hill, NC 27599, USA\\
$^4$Facultad de Ciencias Astron\'omicas y Geof\'isicas (UNLP), La Plata, Argentina\\
$^5$Smithsonian Astrophysical Observatory, 60 Garden St, Cambridge, MA 02138, USA\\
$^6$Dunlap Institute for Astronomy \& Astrophysics, University of Toronto, 50 St. George Street, Toronto M5S 3H4, Canada\\
$^7$Astronomy and Astrophysics Group, Department of Physics, University of Warwick, Gibbet Hill Road, Coventry, CV4 7AL, UK
}

\begin{document}

\maketitle

\begin{abstract}
We present optical high-speed photometry of three millisecond pulsars with low-mass
($<$0.3\,\msun) white dwarf companions, bringing the total number of such systems with
follow-up time-series photometry to five. We confirm the detection of pulsations in one
system, the white dwarf companion to PSR\,J1738+0333, and show that the pulsation
frequencies and amplitudes are variable over many months. A full asteroseismic analysis
for this star is under-constrained, but the mode periods we observe are consistent with
expectations for a $M_{\star} = 0.16 - 0.19 M_{\odot}$ white dwarf, as suggested from
spectroscopy. We also present the empirical boundaries of the instability strip for
low-mass white dwarfs based on the full sample of white dwarfs, and discuss the distinction
between pulsating low-mass white dwarfs and subdwarf A/F stars.
\end{abstract}


\section{Introduction}

White dwarfs (WDs) are the most common type of companion detected around millisecond
pulsars \citep[MSPs,][]{lorimer98}, and thus play a crucial role in establishing the mass
and equation of state of neutron stars. Pulsating WDs can provide a second
clock in these systems, constraining pulsar spin-down and magnetic-field decay (e.g.,
\citealt{kulkarni86}). The characteristic MSP spin-down ages do not necessarily
represent their true ages \citep{tauris12}. Hence, cooling ages of their WD companions provide the only reliable
age measurements in these systems \citep{istrate14}.

Pulsation frequencies and amplitudes depend on the internal structure of the WD,
and thus its evolutionary age \citep{winget08}. We present an observational study of MSPs
with low-mass WD companions with effective temperatures in the regime where
pulsations may be excited by the onset of a surface convection zone.
Hydrogen-dominated WDs with canonical carbon-oxygen cores
($M_{\star}$ $\sim$ 0.6\,\msun) exhibit gravity-mode pulsations between effective
temperatures of $\approx$10,500 and 13,000 K. Detailed asteroseismological analyses
of these stars provide unique constraints on the core carbon-oxygen ratio \citep{giammichele17}, as well as
the thickness of the surface hydrogen and helium layers, which regulate the cooling 
of the star. 

Following the discovery of several extremely low mass (ELM, $M_{\star} \la 0.3 M_{\odot}$)
WDs in the field \citep{kilic10} and as companions to MSPs, \citet{steinfadt10}
predicted that ELM WDs should also pulsate in a similar temperature range
\citep[see also][]{corsico12,vangrootel13}. Their initial search did not find any pulsators
\citep{steinfadt12}, but later searches by \citet{hermes12,hermes13a,hermes13b} found
oscillations in several ELM WDs with pulsation periods ranging from about 20 min
to more than an hour.

However, the instability strip for low-mass WDs is complicated by an overlapping
population of subdwarf A-type (sdA) stars, stars that have spectroscopic surface gravities
comparable to WDs \citep{kepler15, kepler16} but which may be mostly metal-poor
main sequence stars \citep{brown17, pelisoli18}. For example, \citet{bell17} found a 4.3 h
dominant pulsation mode in J1355+1956, a star which has $T_{\rm eff}=8050 \pm 120$ K and
$\log{g}=6.10 \pm 0.06$ based on pure hydrogen atmosphere models. However, this period
significantly exceeds the theoretical limit for pulsations in ELM WDs
\citep{bell17,corsico16}. J1355+1956 is better understood as a SX Phe or $\delta$ Scuti
variable \citep{brown17}.

Given the problems with distinguishing bona fide pulsating low-mass WDs and
pulsating sdA-type stars based on optical spectroscopy, low-mass WD companions to MSPs provide a more reliable 
opportunity to constrain the boundaries of the ZZ Ceti instability strip for these stars.
Because neutron stars are spun up to millisecond periods through accretion in a compact binary
system, MSP companions are expected to be low-mass white dwarfs, and not sdA stars.

There are currently five companions to MSPs with spectroscopic temperatures and surface
gravities within 1500\,K of the extended, low-mass ZZ Ceti instability strip. \citet{kilic15}
discovered pulsations in the WD companion to PSR\,J1738+0333. Here we present the
results from Gemini follow-up photometry of two additional MSP companions, and additional
observations of J1738+0333 from three ground-based facilities. We describe the results for
PSR\,J1911$-$5958A, PSR\,J2234+0611, and PSR\,J1738+0333 in sections 2, 3, and 4, respectively.
We discuss the constraints on the instability strip for low-mass WDs in section 5
and conclude.

\section{PSR\,J1911$-$5958A}

PSR\,J1911$-$5958A is a 3.3 ms pulsar with a $B=22.2$ mag WD
companion in a 0.87 d orbit \citep{damico01}. \citet{bassa06} used optical spectroscopy
of the companion to constrain its parameters, $T_{\rm eff} = 10090 \pm 150$ K and
$\log{g}=6.44 \pm 0.20$. Given the recent improvements in the Stark broadening calculations of
the hydrogen lines in dense plasmas by \citet{tremblay09}, we refit the same spectrum (kindly
provided to us by C. Bassa) with 1D pure hydrogen model atmospheres that include these
improvements \citep{gianninas14}. The best-fit model has $T_{\rm eff} = 10270 \pm 140$ K and
$\log{g}=6.72 \pm 0.05$, and the \citet{tremblay15} 3D model corrections change these
parameters to $T_{\rm eff} = 9980 \pm 140$ K and $\log{g}=6.65 \pm 0.05$.

\citet{bassa06} used $23 \times 600$s and $30 \times 30$s $B-$band acquisition images to check
for optical variability, but found no significant variations. They found an rms scatter of
0.02-0.05 mag for the WD, but this was comparable to the scatter seen in reference
stars of similar brightness. 
\citet{steinfadt12} obtained Hubble Space Telescope observations of PSR\,J1911$-$5958A over
four orbits, but with gaps in the data due to occultations by Earth. They ruled out pulsations 
with amplitudes larger than 16\,mmag.

We obtained time-series photometry of PSR\,J1911$-$5958A using the 8-m Gemini South telescope
with the Gemini Multi-Object Spectrograph (GMOS) on UT 2015 June 17 as part of the queue
program GS-2015A-Q-81. We obtained 88 $\times$ 105 s exposures through an SDSS-$g$ filter over
3.0 h. To reduce the read-out time and telescope overhead to $\approx$15 s, we binned the chip
by 4$\times$4, which resulted in a plate scale of $0.3\arcsec$ pixel$^{-1}$. Observations were 
obtained under thin cirrus with a median seeing of $1.0\arcsec$. We used the standard IRAF Gemini
GMOS routines and the daily bias and sky flats to reduce and calibrate the data, and corrected our times
to the solar-system barycentre using the tools of \citet{eastman10}.
PSR\,J1911$-$5958A has a relatively bright source $3\arcsec$ away. To minimize contamination
from this nearby source, we performed point spread function photometry, and used 15
non-variable reference stars to calibrate the differential photometry. Given the color
differences between the WD and the relatively red reference stars, we fit a second
degree polynomial to the light curve to remove the long term trend due to differential extinction.

\begin{figure}
\vspace{-0.3in}
\includegraphics[width=\columnwidth]{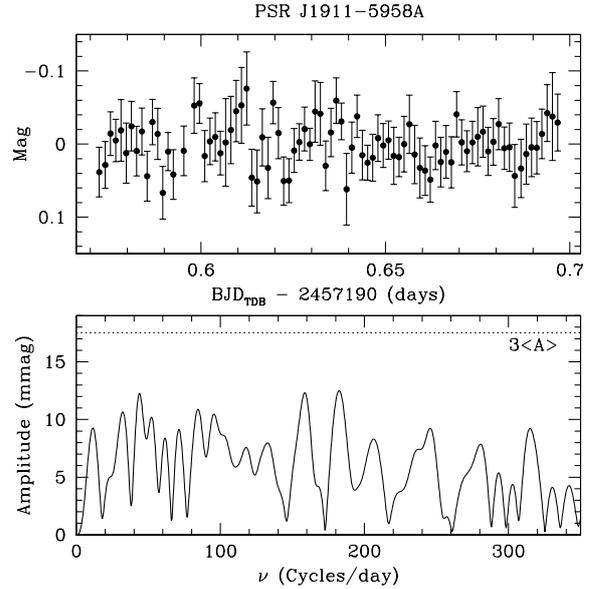}
\vspace{-0.9in}
\caption{The Gemini South light curve (top panel) and its Fourier transform (bottom panel) for
the optical companion to PSR\,J1911$-$5958A. The dashed line marks the 3$\langle {\rm A}\rangle$
significance level (17.4 mmag), as described in the text.}
\label{fig:psr1911}
\end{figure}

Figure \ref{fig:psr1911} shows the Gemini light curve and its Fourier transform for the ELM
WD companion to PSR\,J1911$-$5958A. The median amplitude $\langle {\rm A}\rangle$ in
the Fourier transform is 5.8 mmag, but there are no frequency peaks above 12.5 mmag. Hence, there
is no evidence of pulsations in PSR\,J1911$-$5958A. All but one of the known pulsating ELM WDs
in short period binary systems show pulsations with amplitude larger than this limit.
The exception is SDSS J1112+1117, which displays pulsations with a maximum amplitude of 8.1 mmag
\citep{hermes13a}. Such low level variations would not be detectable for our data on
PSR\,J1911$-$5958A.

\section{PSR\,J2234+0611}

\begin{figure}
\vspace{-0.4in}
\includegraphics[width=\columnwidth]{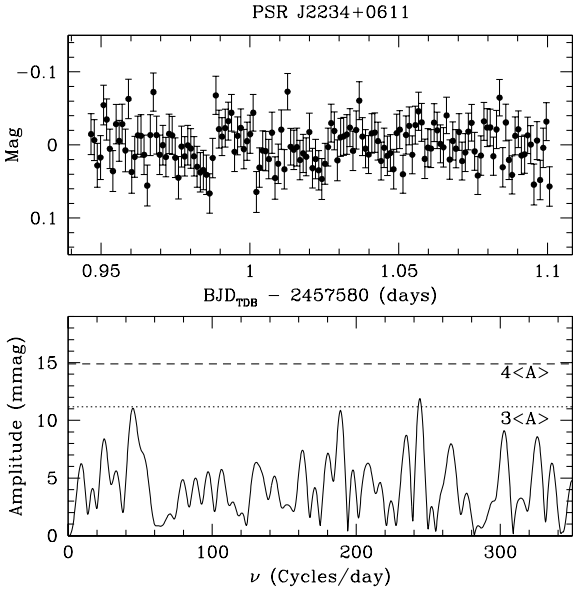}
\vspace{-0.9in}
\caption{The Gemini North light curve (top panel) and its Fourier transform (bottom panel)
for the optical companion to PSR\,J2234+0611. The dotted and dashed lines mark
the 3$\langle {\rm A}\rangle$ and 4$\langle {\rm A}\rangle$ significance levels, respectively.}
\label{fig:psr2234}
\end{figure}

PSR\,J2234+0611 is a 3.6 ms pulsar with a $g=22.2$ mag companion in
an eccentric, $e=0.13$, 32.0 d orbit \citep{deneva13}. \citet{antoniadis16} performed synthetic
photometry of this target by convolving their $26 \times 1420$s follow-up spectra with the $g-$band
filter response curve. They found peak-to-peak variations of about 0.5 mag in the synthetic
photometry, with no obvious correlation with the orbital motion, suggesting that the WD
companion in this system may display high amplitude pulsations.

We obtained time-series photometry of PSR\,J2234+0611 using the same setup as in \S2, but on
the 8-m Gemini North telescope on UT 2016 July 11 as part of the queue program GN-2016B-Q-13.
We obtained 148 $\times$ 75 s $g-$band exposures over 3.7 h. Conditions were photometric with
a median seeing of $0.6\arcsec$. We performed aperture photometry on PSR\,J2234+0611 and seven
nearby reference stars to calibrate the photometry.

Figure \ref{fig:psr2234} shows the Gemini light curve and its Fourier transform for the
companion to PSR\,J2234+0611. There are no significant variations down to a
4$\langle {\rm A}\rangle$ limit of 14.9 mmag for this WD, ruling out pulsations above
this level. Hence, the high amplitude variations seen in the synthetic photometry of
\citet{antoniadis16} were likely not intrinsic to the source, and instead likely caused
by differential refraction effects and/or variable slit-losses.

\section{PSR\,J1738+0333}

Unlike PSR\,J1911$-$5958A and PSR\,J2234+0611, the companion to PSR\,J1738+0333
($V$=21.3 mag) pulsates. Based on 243 $\times$ 50 s exposures
obtained over 5.5 h in 2014, \citet{kilic15} detected three significant periodicities in
the companion to PSR\,J1738+0333 with 10-15 mmag amplitudes. 

To generate a better census of the periods of variability excited in order to complete an
asteroseismic analysis and better constrain the interior structure of this pulsating WD,
we obtained follow-up observations of PSR\,J1738+0333 from three different ground-based
optical facilities. Unfortunately, weather and poor seeing conspired to challenge two of these
datasets.

We were awarded three nights through ESO, from $7-9$ July 2016, to observe PSR\,J1738+0333
with ULTRACAM \citep{dhillon07} mounted as a visitor instrument on the 3.5-m New Technology
Telescope (NTT) in La Silla, Chile. These data were taken simultaneously through
{\textit{u'}},{\textit{g'}},{\textit{r'}} filters and reduced using the ULTRACAM pipeline
software, with standard bias correction and flat-fielding. We performed variable aperture
photometry scaled according to the full width at half maximum and divided our light curve
by two brighter nearby comparison stars. However, we obtained less than 1.5-hr of ULTRACAM
data in cloudy conditions. We binned our 15-s exposures by 6, and our light curve obtained
through the {\textit{g'}} filter had the highest signal-to-noise; still, we did not see any
coherent variability in our ULTRACAM data above a 3$\langle {\rm A}\rangle$ limit of 51 mmag.

We also obtained time-series photometry using the SALTICAM instrument \citep{odonoghue06} on
the 9.2-m South African Large Telescope (SALT) on six visits over five nights: UT 2016 May 3,
June 7, June 9, June 29, and July 4. Given the unique track-length constraints of observing
with SALT, our median observing length in a given night was only 33 min. On 2016 June 7 we
obtained back-to-back tracks for a total of 89 min on target. For each visit we used 45-s
exposures obtained through an SDSS-{\textit{g'}} filter. Seeing for each visit ranged from
$1.8\arcsec$, $1.6\arcsec$, $1.5\arcsec$, $1.5\arcsec$, and $1.8\arcsec$, respectively.

We extracted fixed-aperture photometry from the pipeline-processed SALTICAM data, which is
bias- and flat-field corrected. Only the 2016 June 7 data cover a full cycle of the pulsations
detected from the 2014 Gemini dataset, and so our SALT data are complicated by long-term
atmospheric effects. A Fourier transform of the data from 2016 June 7 show a strong peak at
$1213\pm28$\,s with 23 mmag amplitude, but this peak is not formally significant at
3.4$\langle {\rm A}\rangle$. As with the ULTRACAM data, our SALTICAM photometry does not
well-constrain the long pulsation periods present in PSR\,J1738+0333.

Our most useful data were obtained using Gemini North on UT 2017 May 30, June 2, June 3, and
June 26. We obtained a total of 396 $\times$ 50 s $g-$band exposures. However, there are only
18 observations from June 2, and given the relatively short baseline of the observations, we
exclude those data from our analysis. The other three nights have time baselines of 2.1, 2.5,
and 2.7 h, respectively. Only observations from 2017 May 30 were obtained under photometric
conditions, and the remaining data were obtained under thin cirrus with a median seeing of
$0.6\arcsec$.

\begin{figure}
\vspace{-0.4in}
\includegraphics[width=\columnwidth]{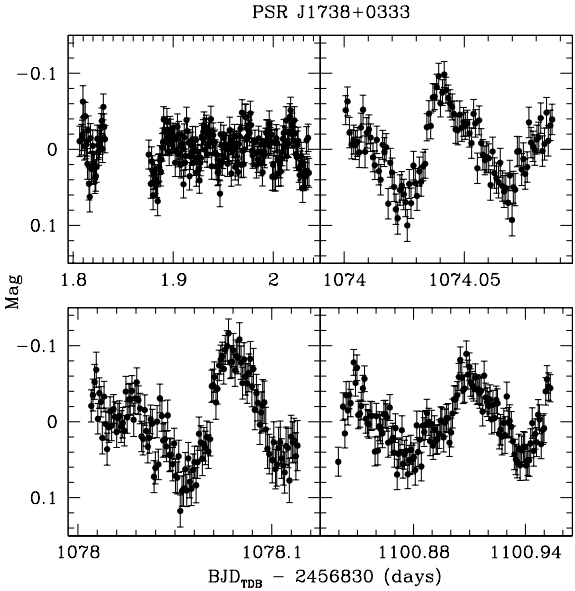}
\vspace{-0.9in}
\caption{The Gemini North light curve for the optical companion to PSR\,J1738+0333 from one
night in 2014 (top left panel) and three nights in 2017.} 
\label{fig:psr1738}
\end{figure}

Figure \ref{fig:psr1738} shows the Gemini light curve of the WD companion to
PSR\,J1738+0333, including the discovery observations from 2014 (top left panel) and the
new data from 2017. At least one of the pulsation amplitudes have significantly increased
in the 2017 data compared to the previous observations. In fact, PSR\,J1738+0333 now shows
0.2\,mag peak-to-peak variations. Interestingly, \citet{antoniadis12} noted $\sim$\,0.05\,mag
scatter in their spectroscopic acquisition images of PSR\,J1738+0333 from 2006. Hence, the
pulsation amplitudes are clearly variable over year and decade timescales.

The coolest carbon-oxygen core pulsating WDs with masses near 0.6\,\msun\ show
amplitude variability; they typically have longer-period pulsations
(e.g., \citealt{kleinman98,hermes14}). Long-baseline {\em Kepler} and {\em K2}
observations have shown that the longest-period modes appear to lose phase coherence
\citep{hermes17}, and a similar phenomenon may be going on at the red edge of the ELM
WD instability strip.

\begin{figure}
\vspace{-0.4in}
\includegraphics[width=\columnwidth]{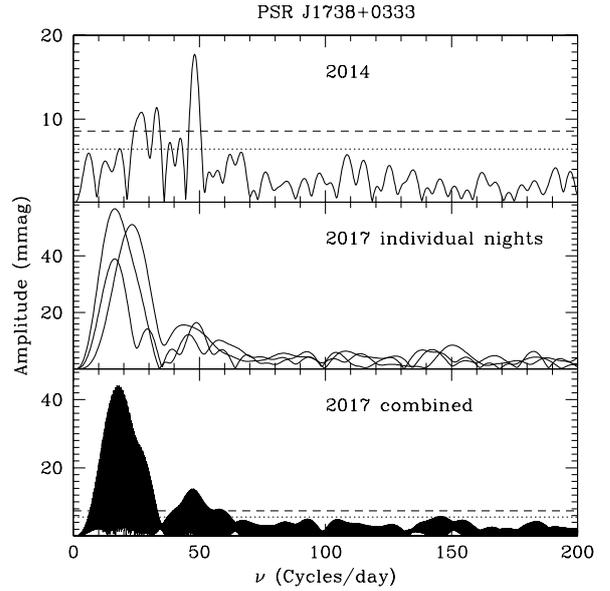}
\vspace{-0.9in}
\caption{Fourier transforms of the optical counterpart to PSR\,J1738+0333 show significant
frequency and/or amplitude variability from 2014 to 2017. This is not unusual for cooler
pulsating WDs of canonical ($\sim$0.6\,\msun), but is the most extreme example
of amplitude variability for a pulsating ELM WD.}
\label{fig:psrft}
\end{figure}

Figure~\ref{fig:psrft} shows the Fourier transforms for the 2014 data (top panel), the
individual nights from the 2017 Gemini dataset (middle panel), and the combined 2017 dataset
along with the 3$\langle {\rm A}\rangle$ and 4$\langle {\rm A}\rangle$ detection limits
(bottom panel). Since each night's data only covers about two cycles of pulsations for the
dominant mode, the peak of the Fourier transform is not well defined for each night.
Combining the data from all three nights of the 2017 dataset, we identify four significant
periodicities above the 4$\langle {\rm A}\rangle$ limit. However, one of these significant
frequencies, 8.76 cycles/day, is identical to our observing window on the night of UT 2017
June 26. Hence, we ignore it in our analysis. We perform 1000 Monte-Carlo simulations to
estimate the uncertainties in frequency and amplitude of each mode. Table 1 presents the
results from this analysis for both the 2014 and 2017 datasets. 

\begin{table}
\centering
\caption{Multi-mode frequency solutions for the WD companion
to PSR\,J1738+0333\label{tab:freq}}
\begin{tabular}{cccccc}
\hline
Dataset & Frequency    & Amplitude  & Period & Model & $\ell$ \\
        & (Cycles/day) &   (mmag)   &  (s)   &  (s)  & \\
\hline
2014  & 27.3919$^{+0.3953}_{-0.4820}$ & 11.1$^{+2.2}_{-1.3}$ & 3154.2 & 3151.8 & 1 \\
\dots & 32.8090$^{+0.5962}_{-0.4881}$ & 10.2$^{+2.1}_{-1.4}$ & 2633.4 & 2632.9  & 1 \\
\dots & 48.3072$^{+0.3338}_{-0.2904}$ & 15.3$^{+2.0}_{-1.3}$ & 1788.6 & 1790.5  & 1 \\
\hline
2017  &  8.7598$^{+0.0422}_{-0.0399}$ & 16.2$^{+1.6}_{-2.8}$  & 9863.2 & \dots & \dots \\
\dots & 17.3475$^{+0.0008}_{-0.0007}$ & 43.5$^{+1.4}_{-3.1}$  & 4980.6 & 4980.2 & 2 \\
\dots & 26.0105$^{+0.0401}_{-0.0405}$ & 21.6$^{+1.2}_{-4.5}$  & 3321.7 & 3323.0 & 1 \\
\dots & 47.1124$^{+0.0021}_{-0.0017}$ & 12.2$^{+1.6}_{-1.8}$  & 1833.9 & 1834.6 & 2 \\
\hline
\end{tabular}
\end{table}

The WD companion to PSR\,J1738+0333 showed 10-15 mmag pulsations with periods
ranging from 1789\,s to 3154\,s in 2014, and 12-44 mmag pulsations with periods ranging
from 1834\,s to 4981\,s in 2017. The pulsation modes at 27.3919 and 26.0105 cycles per day
are consistent in frequency within 3$\sigma$. However, the other modes seem to be
unstable in frequency and/or amplitude. 

We attempted asteroseismology of this star by fitting the observed periodicities with a new
set of low mass WD models that include a range of hydrogen envelope thicknesses.
\citet{calcaferro18} present the details of these models for 0.15-0.44 $M_{\odot}$ WDs
with hydrogen envelope masses of $10^{-5.8}$ to $10^{-1.7} M_{\star}$. We use the merit
function $\chi^2$ as defined by equation 2 of \citet{calcaferro17} to find the best-fit model.
Given the unstable periods and amplitudes, we treat the 2014 and 2017 datasets separately.

Table 1 presents the best-fit model periods (the last column) for each dataset assuming that
all of the observed periods correspond to $g$ modes with $\ell=1$ or 2. The 2014 dataset is
best-explained by a model that has $M_{\star} = 0.192 M_{\odot}$,
$M_{\rm H} = 9.5 \times 10^{-4} M_{\star}$, $T_{\rm eff} = 9273$ K, $\log{g}=6.63$, and
three $\ell=1$ pulsation modes at 1790.5, 2632.9, and 3151.8 s. The average difference between
the observed and predicted periods in this model is $\Delta P = 1.6$ s. On the other hand, the
2017 dataset is best-explained by a model that has 
$M_{\star} = 0.161 M_{\odot}$, $M_{\rm H} = 1.7 \times 10^{-2} M_{\star}$, $T_{\rm eff} = 8883$ K,
and $\log{g}=6.05$, with a mixture of $\ell=1$ and 2 modes and an average difference in period
of $\Delta P = 0.8$ s. This model implies a WD cooling age of 3 Gyr.

PSR\,J1738+0333's WD companion has 3D corrected spectroscopic values of
$T_{\rm eff}= 8910 \pm 150$ K and $\log{g}=6.30 \pm 0.10$ \citep{tremblay15}, and an
independent estimate of $\log{g}=6.45 \pm 0.07$ based on the orbital period decay, mass ratio,
parallax, and absolute photometry of the system \citep{antoniadis12}. Even though the asteroseismological fits confirm
the nature of the companion to PSR\,J1738+0333 as a low-mass WD, the best-fit model
depends heavily on the exact values of the periods. For example, the 3321.7 s mode in the 2017
dataset is uncertain by about 10 s. Repeating our asteroseismological analysis of this dataset
by taking into account the period uncertainties, we find best-fit solutions with
$M_{\star} = 0.155-0.192 M_{\odot}$, 
$M_{\rm H} = 4 \times 10^{-6} M_{\star}$ to $4 \times 10^{-3} M_{\star}$, and cooling ages
ranging from 0.03 to 1.4 Gyr. This exercise shows that our asteroseismic constraints are not
robust, because there are only three pulsation modes detected in each dataset with relatively
large errors. Combining the results from both the 2014 and 2017 data, we conclude that
PSR\,J1738+0333's companion has $M_{\star} = 0.155-0.192 M_{\odot}$, $T_{\rm eff} = 8840-9270$ K,
a poorly constrained hydrogen layer mass of 
$M_{\rm H}/M_{\star} = 4 \times 10^{-6}$ to $1.7 \times 10^{-2}$, and a WD cooling
age of 0.03-3 Gyrs.

Gravity-mode pulsations can be driven by tidal excitations in compact binary white dwarf systems \citep{fuller11}. 
Tidally-forced oscillations are typically seen in eccentric binaries and with flux variations at integer multiples
of the orbital frequency \citep{fuller12}, which is 2.82 cycles per day for PSR J1738+0333. Due to the relatively
large errors, any of the observed frequencies in the 2014 dataset can be explained by tidal excitation. However,
the eccentricity of the PSR J1738+0333 binary is very low, $3.5 \pm 1.1 \times 10^{-7}$ \citep{antoniadis12}, and
none of the pulsation frequencies in the 2017 dataset are multiples of the orbital frequency. Hence, tidal excitation
is unlikely to explain the pulsations seen in PSR J1738+0333's WD companion.

\section{The ZZ Ceti Instability Strip for Low-Mass WDs}

There are five MSP + ELM WD system where optical spectroscopy puts the WD
within 1500\,K of the instability strip for low-mass WDs. We now have high-cadence
photometry for all five systems, and only PSR\,J1738+0333's companion pulsates to
detectable amplitudes. Table 2 presents the physical parameters for these five systems
based on the 1D model atmosphere fits and the 3D corrections of \citet{tremblay15}. 

The companion to PSR\,J1012+5307 has two different $\log{g}$ estimates in the literature,
$6.75 \pm 0.07$ \citep{vankerkwijk96} and $6.34 \pm 0.20$ \citep{callanan98}, and both are based on
an analysis that does not include the recent Stark broadening profiles of \citet{tremblay09}.
To improve the constraints on its physical parameters, we obtained four back-to-back $12-15$
min spectra of PSR\,J1012+5307 using the MMT Blue Channel Spectrograph equipped with the
832 lines mm$^{-1}$ grating and the $1\arcsec$ slit on UT 2016 Dec 2. This setup provided
spectra with 1\AA\ resolution over 3600-4500 \AA. After shifting each spectrum to rest velocity,
we fitted the summed spectrum using 1D pure hydrogen model atmospheres that include the
improved Stark broadening profiles \citep{gianninas14}. The best-fit model has
$T_{\rm eff} = 8630 \pm 120$ K and $\log{g}= 6.63 \pm 0.09$, and the 3D model corrections lower
these values to $T_{\rm eff} = 8500$ K and $\log{g}= 6.31$, respectively.

\begin{table}
\centering
\caption{3D model corrected parameters of the WD companions to five MSPs.}
\begin{tabular}{llcc}
\hline
Object & $T_{\rm eff}$ & $\log{g}$    &   Source  \\
 PSR\,  &      (K)     & (cm s$^{-2}$) &           \\
\hline
J1012+5307    & $8500 \pm 120$ & $6.31 \pm 0.09$ & This paper \\
J1738+0333    & $8910 \pm 150$ & $6.30 \pm 0.10$ & \citet{antoniadis12} \\
J1909$-$3744  & $8920 \pm 150$ & $6.81 \pm 0.15$ & \citet{antoniadis13} \\
J1911$-$5958A & $9980 \pm 140$ & $6.65 \pm 0.05$ & This paper \\
J2234+0611    & $8600 \pm 190$ & $6.97 \pm 0.22$ & \citet{antoniadis16} \\ 
\hline
\end{tabular}
\end{table}

Looking at Table 2, the only pulsating companion to a MSP (J1738+0333) has $T_{\rm eff}=8910$ K
and $\log{g}=6.3$. The companion to PSR\,J1909$-$3744 has an identical temperature to J1738+0333,
but its surface gravity is 0.5 dex higher. Similarly, PSR\,J1012+5307's companion has an
identical surface gravity to J1738+0333, but its temperature is about 400 K cooler. MSP
companions cooler than or more massive than J1738+0333 do not pulsate, and we suspect this
pulsating WD is near the cool (red) edge of the extended ZZ Ceti instability strip.
As WDs move through the instability strip their convection zones deepen, driving
longer-period pulsations. These long-period pulsations also tend to be the least stable in
frequency and amplitude.

\begin{figure}
\vspace{-0.4in}
\hspace{-0.2in}\includegraphics[width=3.7in]{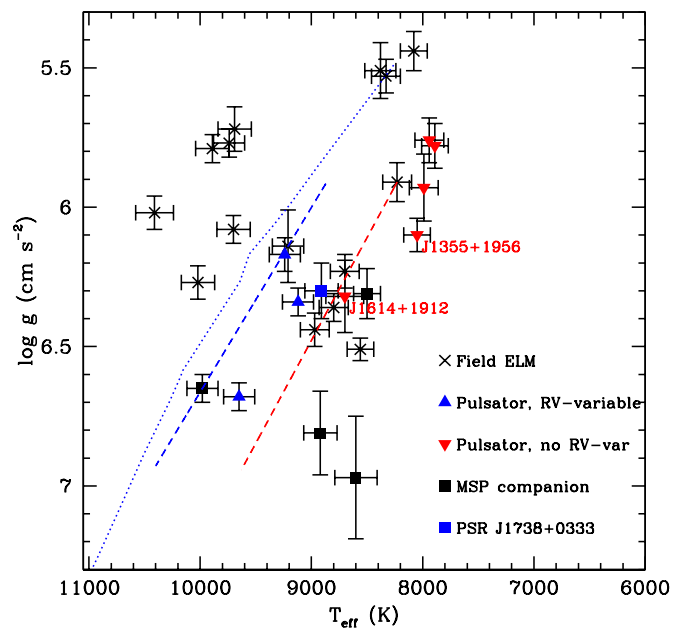}
\vspace{-0.9in}
\caption{Temperatures and surface gravities (3D corrected) for low-mass WDs with
follow-up time-series photometry. Filled squares represent companions to MSPs.
Blue  symbols mark the pulsating stars that show high amplitude radial velocity (RV) variations
since they are in short period binary systems, while red symbols mark the pulsating stars that
are not RV variable. The dotted line shows the theoretical blue edge of the instability strip from
\citet{corsico16}, and the dashed lines mark our empirical boundaries.}
\label{fig:strip}
\end{figure}

Figure~\ref{fig:strip} shows the physical parameters of these five MSP + WD systems
(filled squares) plus the field ELM WDs with follow-up time-series photometry and
spectroscopy. There are four pulsating ELM WDs in short-period binaries and those are
marked with blue symbols. These are J1112+1117, J1518+0658, J1840+6423, and
PSR J1738+0333's WD companion. The first three have $>5\sigma$ significant parallaxes (1.2-2.8 mas)
in Gaia Data Release 2 \citep{gaia18}, which confirm them as WDs with absolute Gaia G-band
magnitudes of $M_{\rm G} =$ 8.4-9.8. The latter has parallax measurements in the radio \citep{antoniadis12}. These four pulsators occupy a similar parameter space and there is no question 
about their nature; they are clearly WDs. 

Red symbols show the other five pulsating ELM candidates that are not in short period binary systems.
\citet{brown17} demonstrate that unlike the published ELM WD binaries, the majority of
the sdA-like objects show no evidence for short-period or high-amplitude radial velocity variability.
Interestingly, four of these known pulsators are significantly cooler than J1738+0333 and they
have $T_{\rm eff}\approx 8000$ K. In addition, \citet{bell17} found a dominant 4.3 hr pulsation
period in one of these stars, J1355+1956, which likely rules out pulsations from a WD
and instead favors a high-amplitude $\delta$ Scuti star. The remaining four stars, J1614+1912,
J1735+2134, J2139+2227, and J2228+3623 also have Gaia parallax measurements (0.18-0.26 mas),
which correspond to $M_{\rm G} =$ 2.4-3.9 mag. These four objects are too bright to be WDs.

Ignoring these objects, the dashed lines show our empirical boundaries of the instability strip based
on the field and MSP + ELM WD samples. The boundaries are

\begin{equation}
\begin{aligned}
(\log g)_{\rm blue} = 6.6234 \times 10^{-4} (T_{\rm eff}) + 0.03987 \\
(\log g)_{\rm red} = 7.3754 \times 10^{-4} (T_{\rm eff}) - 0.15987
\end{aligned}
\end{equation}

Given the relatively small number of objects in this figure, these boundaries are preliminary.
Nevertheless, our empirical blue edge is consistent with the theoretical predictions from
\citet[][shown as a dotted line in Figure~\ref{fig:strip}]{corsico16}.

\section{Conclusions}

We present high-speed photometry of three MSP + ELM WD systems and find pulsations only
in the companion to PSR\,J1738+0333. We find that the observed modes have variable periods
and amplitudes on a timescale of weeks to years. Our asteroseismic analysis constrains the
stellar mass to 0.16-0.19 $M_{\odot}$ but provides limited constraints on the surface hydrogen
layer mass due to the small number of modes observed and the relatively large errors in the
measured periods and the paucity of modes observed.

We compare the physical parameters of the MSP + WD systems with the pulsating field
ELM WDs. We find two sets of objects, bona fide pulsating WDs with
temperatures near 9,000 K and a second set of pulsating stars with temperatures near 8,000 K
that are likely sdA stars. We use the current sample of pulsating and non-pulsating ELM
WDs to constrain the boundaries of the instability strip. Our empirical blue edge
is consistent with theoretical predictions, but high speed photometry of additional low-mass
WDs with $T_{\rm eff}=$ 9000-10,000 K would be helpful for improving these boundaries.

\section*{Acknowledgements}

We gratefully acknowledge support from the NASA grant NNX14AF65G. Support for this work was
provided by NASA through Hubble Fellowship grant \#HST-HF2-51357.001-A, awarded by the
Space Telescope Science Institute, which is operated by the Association of Universities for
Research in Astronomy, Incorporated, under NASA contract NAS5-26555. MJG acknowledges funding
from an STFC studentship via grant ST/N504506/1. MK thanks the University of Edinburgh's Institute
for Astronomy Wide Field Astronomy Unit staff for their hospitality during his sabbatical visit.

Some of the observations reported in this paper were obtained with the Southern African Large
Telescope under programme 2016-1-SCI-017, as well as observations made with ESO Telescopes at
the La Silla Observatory under programme ID 097.D-0342.

\end{document}